\documentclass[prl, aps, english, twocolumn, hyperref,superscriptaddress,showpacs]{revtex4}
\usepackage{graphicx}
\usepackage{amssymb, amsmath, amsfonts, color, rotating, multirow, graphicx, bm}
\usepackage[dvipdfm,bookmarks=false,pdfstartview=FitH,hyperindex=true, colorlinks, linkcolor=blue, citecolor=blue]{hyperref}

\begin{document}
\author{Ran Wei}
\affiliation{Hefei National Laboratory for Physical Sciences at Microscale and Department
of Modern Physics, University of Science and Technology of China, Hefei, Anhui
230026, China}
\affiliation{Laboratory of Atomic and Solid State Physics, Cornell University, Ithaca, NY, 14850}
\author{Erich J. Mueller}
\affiliation{Laboratory of Atomic and Solid State Physics, Cornell University, Ithaca, NY, 14850}
\title{Pair density waves and vortices in an elongated two-component Fermi gas}
\date{\today}


\begin{abstract}
We study the vortex structures of a two-component Fermi gas experiencing a uniform effective magnetic field in an anisotropic trap that
interpolates between quasi-one dimensional (1D) and quasi-two dimensional (2D). At a fixed chemical potential, reducing the anisotropy
(or equivalently increasing the attractive interactions or increasing the magnetic field) leads to instabilities towards
pair density waves, and vortex lattices. Reducing the chemical potential stabilizes the system.
We calculate the phase diagram, and explore the density and pair density. The structures are similar to those
predicted for superfluid Bose gases. We further calculate the paired fraction, showing
how it depends on chemical potential and anisotropy.
\end{abstract}
\maketitle

\textit{Introduction ---}
Quantized vortices play an essential role in understanding the behavior of type-II superconductors and superfluids such as $^{3}$He.
In cold gases, these vortices were the smoking gun for superfluidity \cite{Zwierlein2005}.
Here we study how confinement influences the vortex structures in a trapped  gas of ultracold fermions.
We use the microscopic Bogoliubov-de-Gennes (BdG) equations, and consider
anisotropic traps that interpolate between quasi-one dimensional (1D) and quasi-two dimensional (2D).

The behavior of topological defects in confined geometries can be quite rich. A good example is rotating
bosons in anisotropic traps \cite{Sinha2005}, where one sees multiple transitions in the structure of vortex
lattices as the parameters are changed. Most intriguing, in the quasi-1D limit one sees a ``roton" spectrum which softens
as the rotation rate increases, signaling an instability to form a snake-like density wave.
With recent experimental developments \cite{Lin2009nature}, we expect these structures can soon be
explored in Bose gases, and related studies will be undertaken in Fermi gases.
In the Fermi gas, we find parallels to all of the predicted boson physics.
The single particle instability which drives density waves in the Bose case becomes a collective
instability for the fermions, and instead drives pair density waves \cite{agterberg2008}. For a range of parameters we
even find that the order parameter has the form predicted by Larkin and Ovchinnikov \cite{LO1965}
for a polarized gas.

In very different contexts, studies of vortices in confined geometries
lead to a number of interesting and important results such as
``non-Hermitian" quantum mechanical analogies \cite{confinedvortices},
and the destruction of superconductivity via phase slips \cite{superconductingwire}.
Generically, reducing the dimensionality enhances fluctuations, leading to novel effects.

Driven partially by increased computer power and partially by interest in the BCS-BEC crossover,
a number of research groups have recently produced Bogoliubov-de-Gennes (BdG) or density functional calculations
of single vortices \cite{singlevortex}, and vortex lattices \cite{vortexlattice}.
These have largely been 2D or three dimensional (3D) calculations, with translational symmetry along
the magnetic field. The numerical challenges of these calculations come from the large basis set needed to
describe the single particle states. By truncating to the lowest Landau level,
one can greatly simplify the problem \cite{LLLBdG}.
As we explain below this limit is experimentally relevant \cite{experimentLLL,Aidelsburger2011}.

\textit{Model --- }
We start from the Hamiltonian of a spin balanced two-component Fermi gas, with
a total number of particles $N=\int(\Psi_{\uparrow}^\dagger\Psi_{\uparrow}+
\Psi_{\downarrow}^\dagger\Psi_{\downarrow})d\bm{r}$ and chemical potential $\tilde\mu$,
\begin{eqnarray}
\mathcal{K}=\int\biggl(\sum_{\sigma=\uparrow,\downarrow}\Psi_{\sigma}^\dagger
H_0\Psi_{\sigma}+H_{\rm int}\biggr) d\bm{r}-\tilde\mu N,
\end{eqnarray}
where the single particle Hamiltonian
$H_0=(p_x-By)^2/2m + p_y^2/2m + p_z^2/2m + V(\bm{r})$,
describes a neutral atom of mass $m$ and momentum $\bm{p}$
experiencing a uniform effective magnetic field $B$ in the $z$ direction (Landau gauge),
where the harmonic trap is $V(\bm{r})=\frac{1}{2}m(\omega_y^2y^2+\omega_z^2z^2)$,
and the inter-component interaction $H_{\rm int}=-g\Psi_{\uparrow}^\dagger\Psi_{\downarrow}^\dagger\Psi_{\downarrow}\Psi_{\uparrow}$,
is attractive, with the coupling constant related to $s$-wave scattering lengths
$a_s$ via $g=-4\pi\hbar^2 a_s/m$ ($g>0$) \cite{regulation}.
We do not treat the case where $g<0$, in which the physics is more involved \cite{Zwerger2004}.
This single particle Hamiltonian is readily engineered in cold atoms either by using two counter-propagating Raman beams
with spatially dependent detuning \cite{Lin2009nature} or rotating the
gas in anisotropic traps where the rotation rate approaches the weakest trapping frequency \cite{dalibard2002}.
When $\omega_z$ is large, this model can be tuned from quasi-1D to quasi-2D by changing $\omega_y$.

\textit{(A) lowest Landau level --- } Following Sinha \textit{et.al} \cite{Sinha2005},
the single particle Hamiltonian is readily diagonalized, with
eigenstates labeled by three quantum numbers $K,n,n^\prime$, and energies given by
\begin{eqnarray}
E_{nn^\prime}(K)={\cal E} K^2 + n \hbar\omega_z+n^\prime \hbar \tilde \omega_c,
\end{eqnarray}
where the effective cyclotron frequency is $\tilde \omega_c=\sqrt{\omega_y^2+\omega_c^2}$,
the cyclotron frequency is $\omega_c=B/m$, the characteristic energy of motion in the $x$ direction is
${\cal E}=\hbar\omega_y^2/4\tilde\omega_c$, and we have neglected the zero-point energy.
The dimensionless wave-number $K=\sqrt{2}\tilde\ell k$ labels the momentum $k$ along the $x$ direction,
where the effective magnetic length is $\tilde\ell=\sqrt{\hbar/m\tilde\omega_c}$.
The discrete quantum numbers $n$ and $n^\prime$ corresponds to the number of nodes in the $z$ and $y$ directions.
In the absence of confinement in the $y$ direction, ${\cal E}\rightarrow0$, and we recover degenerate Landau levels.
Hence, we refer to $n$ as the Landau level index.
If the interaction energy per particle $\langle H_{\rm int}/N\rangle$
and the characteristic ``kinetic energy" $\langle {\cal E} K^2\rangle$ are small compared to $\hbar \tilde\omega_c$ and $\hbar\omega_z$,
one can truncate to the lowest eigenstates with $n=n^\prime=0$, which are of the form
\begin{equation}
\label{sp}
\phi_K(\bm{r})=\frac{1}{\sqrt{\pi\tilde\ell d_zL}}e^{i\frac{Kx}{\sqrt{2}\tilde\ell}}e^{-\frac{(y-y_K)^2}{2\tilde\ell^2}}e^{-\frac{z^2}{2d_z^2}},
\end{equation}
where $y_K=\sqrt{2}\omega_cK\tilde\ell/2\tilde\omega_c$, $d_z=\sqrt{\hbar/m\omega_z}$ and
$L$ is the length in the $x$ direction.

The conditions allowing us to truncate to the lowest Landau level constrain the 3D density $n_{3D}$ and magnetic field strength $B$.
For example, the condition $\langle H_{\rm int}/N\rangle\ll \hbar\tilde\omega_c\sim\hbar\omega_z$
requires $n_{3D}\ll \hbar\tilde\omega_c/g\sim\hbar\omega_z/g$.
The other condition, ${\cal E} K^2 \ll \hbar \tilde \omega_c$, requires $B\gg m\omega_y$.
While such fields are challenging to produce in cold atoms, they are
not completely unreasonable. In a very recent experiment performed by I.Bloch's group \cite{Aidelsburger2011},
the density is $n_{3D}\sim10^{13}$cm$^{-3}$, and the cyclotron frequency is $\omega_c\sim100$kHz.
Since this experiment involves coupled ``wires",  it is natural to use them for quasi-1D studies.
Note, the magnetic field is ``staggered" in that experiment, while we consider the uniform case.

Letting $a_K$ annihilate the state in Eq.~(\ref{sp}), one has an effective 1D model,
\begin{eqnarray}
\label{LLL}
H/{\cal E}=\sum_{K,\sigma}(K^2-\mu)a_{K\sigma}^\dagger a_{K\sigma}
+\beta\sum_{q}f^\dagger(q)f(q),
\end{eqnarray}
where $f(q)\equiv\sum_Ke^{-1/8(2K-q)^2}a_{q-K\downarrow}a_{K\uparrow}$,
the dimensionless chemical potential is $\mu=\tilde\mu/{\cal E}$
and the effective interaction parameter is
$\beta=-\frac{2mg}{\pi\hbar^{2}L}(\frac{\omega_z}{\tilde\omega_c})^{1/2}(\frac{\tilde\omega_c}{\omega_y})^2$.
From the definition of $\beta$, one sees that increasing the interaction strength $g$
has the same effect as increasing the magnetic field $B$,
increasing the $z$-confinement $\omega_z$, or reducing the $y$-confinement $\omega_y$.
In the following, we will investigate the properties of the confined Fermi gas by studying Eq.(\ref{LLL}).
One can show that the interaction in Eq.(\ref{LLL}) is equivalent to
$\beta\sum_qf^\dagger(q)f(q)=\beta\int d\bm{r}F^\dagger(\bm{r})F(\bm{r})$,
where $F(\bm{r})=\sum_qf(q)\phi_q(\bm{r})$.

\begin{figure}[!htb]
\includegraphics[width=7cm]{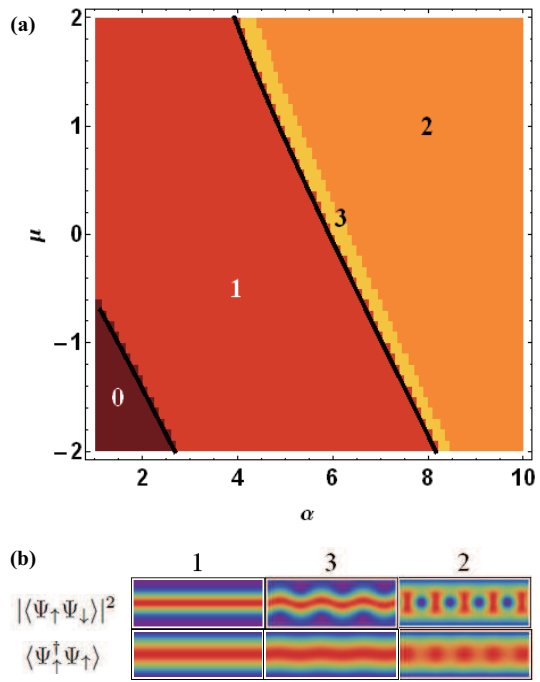}
\caption{(color online) (a): The structure of phase diagram as a function of $\alpha$ and $\mu$. The value of $\xi$
(the number of nonzero $\Delta_{|n|K_0}$) is denoted in each region.
The two black solid curves are the boundaries of two continuous transitions:
$\xi=0\leftrightarrow\xi=1$ and $\xi=1\leftrightarrow\xi=3$. They show a fairly good agreement with numerics.
(b): The structures of pair density $|\langle\Psi_\uparrow\Psi_\downarrow\rangle|^2$
and density $\langle\Psi_\uparrow^\dagger\Psi_\uparrow\rangle$ in the corresponding regions.
The color key is shown in Fig.\ref{lattice}.}
\label{pd}
\end{figure}

\begin{figure}[!htb]
\includegraphics[width=8cm]{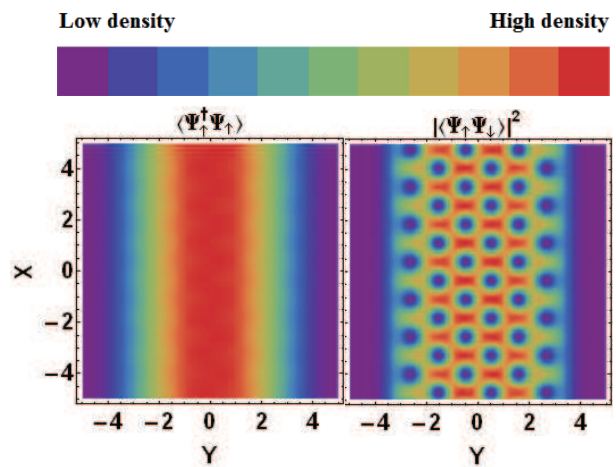}
\caption{(color online) The profile of density (left panel) and pair density (right panel)
at $\alpha=65,\mu=2$, where the dimensionless coordinates are
${\rm X}=x/\sqrt{2}\tilde\ell,{\rm Y}=y/\sqrt{2}\tilde\ell$. The color key is shown on the top.}
\label{lattice}
\end{figure}

\textit{(B) Bogoliubov de Gennes approach --- }
We introduce the pair field $\Delta_q=\beta\langle f(q)\rangle$, and its transform
$\Delta(\bm{r})=\beta\langle F(\bm{r})\rangle$.
We neglect the fluctuation $(f^\dagger(q)-\Delta_q^*/\beta)(f(q)-\Delta_q/\beta)$ to
reduce Eq.(\ref{LLL}) to a bilinear form,
\begin{eqnarray}
\notag H/{\cal E}&=&\sum_{K,\sigma}(K^2-\mu)a_{K\sigma}^\dagger a_{K\sigma}\\
&+&\sum_{q}\left(\Delta_q^*f(q)+\Delta_qf^\dagger(q)-|\Delta_q|^2/\beta\right).
\end{eqnarray}
Given $\Delta_q$, one can diagonalize $H$, and then impose self-consistency.
For arbitrary $\Delta_q$, this process is unwieldy \cite{vortexlattice}.
We here introduce two approximations which make the numerical calculations more efficient.
First, we assume $\Delta_q$ is non-vanishing only when the central momentum of the
paired fermions is $q=nK_0$, where $n=0,\pm1,\pm2,...$. The characteristic
wave-number $K_0$ is taken to be a variational parameter. This is equivalent to assuming
$\Delta(\bm{r})$ is periodic in the $x$ direction and treating the wavelength variationally.
Second, we restrict ourselves to consider the symmetric pair field: $\Delta_{q}=\Delta_{-q}$.
This implies a spatially symmetric field $\Delta(\bm{r})=\Delta(-\bm{r})$.
Under these assumptions, the Hamiltonian is reduced to

\begin{eqnarray}
\label{sime}
\notag H/{\cal E}&=&\sum_{K,\sigma}(K^2-\mu)a_{K\sigma}^\dagger a_{K\sigma}-\sum_{n}|\Delta_{|n|K_0}|^2/\beta\\
&+&\sum_{n}\left(\Delta_{|n|K_0}^*f(nK_0)+\Delta_{|n|K_0}f^\dagger(nK_0)\right).
\end{eqnarray}
Since Eq.(\ref{sime}) will be calculated by taking the continuum limit
$\sum_{K}\rightarrow(\sqrt{2}L/4\pi\tilde\ell)\int dK$ (\textit{see Supplemental materials}),
it is useful to introduce a positive parameter $\alpha=-\sqrt{2}L\beta/4\pi\tilde\ell$
to characterize the effective attractive interaction.
For small $\alpha$, we find $\Delta_{|n|K_0}\neq0$ for only a few values of $n$.
We define $\xi$ to be the number of nonzero $\Delta_{|n|K_0}$.
The various phases can be distinguished by looking at the pair density $|\langle\Psi_\uparrow\Psi_\downarrow\rangle|^2$
and/or the particle density $\langle\Psi_\uparrow^\dagger\Psi_\uparrow\rangle$ (see Fig.\ref{pd}(b)).
The features are clearest in the pair density.
If more than one $\Delta_{nK_0}$ is nonzero, we have either a pair density wave or vortices.
For example, the case $\xi=3$ ($\Delta_0\neq0,\Delta_{\pm K_0}\neq0$), as illustrated in Fig.\ref{pd}(b),
corresponds to a pair density wave where
$|\langle\Psi_\uparrow\Psi_\downarrow\rangle|^2$ has corrugations. The case $\xi=2$
($\Delta_0=0,\Delta_{\pm K_0}\neq0$), consists of a single row of vortices. Larger $\xi$,
for example in Fig.\ref{lattice}, corresponds to a vortex lattice.
The case $\xi=2$ gives an order parameter which can formally be identified with the
Larkin-Ovchinnikov (LO) state \cite{LO1965} (see also \cite{FF1964}). Here, $\Delta_K$ is nonzero
except when $K=\pm K_0$. Defining an effective 1D order parameter $\Delta^{1D}(x)=\sum_Ke^{iKx}\Delta_K$,
we have $\Delta^{1D}(x)=2\Delta_{K_0}\cos K_0x$. Note that unlike the LO state, the physical order
parameter $\Delta(\bm{r})=\sum_K\Delta_K\phi_K(\bm{r})$, is not a simple cosine. Also note that
unlike LO's model, here we assume both spin states have equal chemical potentials.
Instead of being driven by the polarization, our instability towards a paired density
wave is driven by the form of the effective 1D interaction.

When $\xi=1$ ($\Delta_0\neq0$), Eq.(\ref{sime}) can be analyzed analytically (\textit{see Supplemental materials -- A}).
One readily obtains the gap equation,
\begin{eqnarray}
\label{gap}
\frac{1}{\alpha}&=&\int\frac{e^{-K^2}}{2\epsilon_K}dK,
\end{eqnarray}
and the number equation,
\begin{eqnarray}
\label{number}
N&=&\frac{\sqrt{2}L}{4\pi\tilde\ell}\int(1-\frac{\epsilon_0}{\epsilon_K})dK.
\end{eqnarray}
where $\epsilon_{K}=\sqrt{\epsilon_0^2+|\Delta_0|^2e^{-K^2}}$ and $\epsilon_0=K^2-\mu$.

Unlike the traditional case, the integrand in the RHS of Eq.(\ref{gap}) has a factor $e^{-K^2}$ in the numerator,
which dominates the behavior of the integrand for $K\gg1$.
If $\mu\gg1$ (meaning in physics units $\tilde\mu\gg{\cal E}$), and
$\Delta_0$ is sufficiently small, the integrand in Eq.(\ref{gap}) is bimodal.
There is a gentle peak of height $\frac{1}{2\mu}$ and width $1$ centered at $K=0$,
and a sharp peak of height $\frac{e^{-\mu/2}}{2|\Delta_0|}$ and width $\frac{|\Delta_0|e^{-\mu/2}}{\sqrt{\mu}}$
centered at $K=\sqrt{\mu}$. The power-law tails of this sharp peak give a contribution to the integral
which scales as $A\frac{e^{-\mu}}{\sqrt{\mu}}\log|\Delta_0|$ as $\Delta_0\rightarrow0$, where $A$ is a constant.
Solving Eq.(\ref{gap}) in this regime yields an extremely small order parameter.
In this weak pairing limit, our numerics are unstable and the vortex lattices are better treated by expanding the energies
in power of $\Delta_0$ \cite{abrikosov1957}.

Another instructive limit is $\mu<0$ and $N/L\rightarrow0$,
where the behavior is dominated by two-body physics.
Eq.(\ref{gap}) then becomes the Schr\"{o}dinger equation of a two-body problem
in momentum space \cite{Salpeter1951},
i.e., $\alpha=2/\int e^{-K^2}/({K^2-\mu})dK$,
where the two-body binding energy $\eta$
is identified with twice the chemical potential, $\eta=2\mu(\alpha)$.

\textit{Phase diagram --- }
We numerically minimize the energy by studying Eq.(\ref{sime}) (\textit{see Supplemental materials -- B}).
We find discrete jumps in $\xi$ as a function of the dimensionless attractive interaction
$\alpha$ and the dimensionless chemical potential $\mu$.
The resulting phase diagram is shown in Fig.\ref{pd}(a).
The darkest red region ($\xi=0$) is the vacuum with no particles.
Increasing $\alpha$ and/or $\mu$ brings one to a quasi-1D superfluid state. This state,
characterized by $\xi=1$, has no vortices and is translational invariant in
the $x$ direction. The $\xi=0$ to $\xi=1$ transition is continuous with $\Delta_0\rightarrow0$
and $N/L\rightarrow0$ at the boundary. Further increasing $\alpha$ and/or $\mu$ leads to an
instability towards a $\xi=3$ state (the narrow yellow region).
This state breaks translational symmetry. The transition is continuous, and
the boundary can be found via a linear stability analysis of the $\xi=1$ state (\textit{see Supplemental materials -- C}).
At larger $\alpha$ and/or $\mu$, there is a
discontinuous transition to a state with $\xi=2$. This sequence of instabilities closely
mirrors what is found in calculations for Bose gases \cite{Sinha2005}.

\begin{figure}[!htb]
\includegraphics[width=8cm]{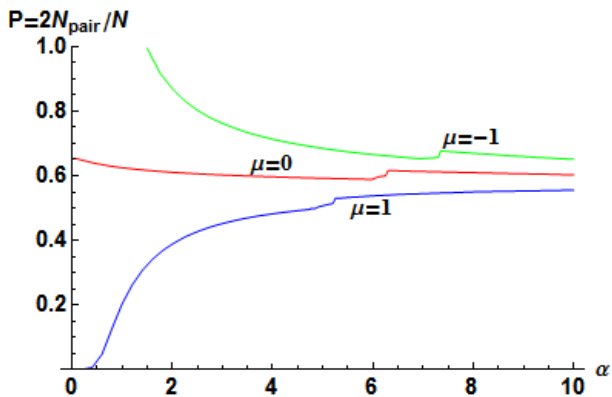}
\caption{(color online)
The pair fraction $P=2N_{\rm pair}/N$ versus $\alpha$ with $\mu=-1,0,1$.
The exponential small $P$ for $\mu=1$ at $\alpha\rightarrow0$ is reminiscent of the BCS limit,
and the large value of $P$ for $\mu=-1$ at $\alpha\approx1.5$ is analogous to the BEC limit.
The kink on each curve corresponds to the $\xi=3\leftrightarrow\xi=2$ phase transition.}
\label{fraction}
\end{figure}

\textit{Pair fraction ---}
It is useful to put these results in the context of the BCS-BEC crossover. In 3D Fermi gases one thinks of
the superfluid with $\mu<0$ as being formed from tightly bound bosonic pairs, analogous to $^4$He.
The superfluid with $\mu>0$ is instead thought of within a BCS picture where diffuse pairs are formed by atoms
at the Fermi surface. One can continuously tune between these two idealized limits by taking $\mu$ through zero:
the size of the pairs varies continuously.
Our approach to gaining insight into analogies with the 3D BCS-BEC crossover is to study the
pair fraction $P=2N_{\rm pair}/N$ \cite{pair}, as in Fig.\ref{fraction}.
While some of the qualitative features of the 3D crossover persist in our effective 1D model, many of the details differ.

To understand this figure, one must note that in a quasi-1D system
the ratio of the interaction to the kinetic energy is inverse proportional to
the density, thus the strongly interacting regime can be reached by making the density small, or by making
$\alpha$ large. The density increases monotonically with $\mu$, but varies in a more complicated fashion with $\alpha$.
For small $\alpha$ and $\mu>0$ we find $\partial N/\partial\alpha<0$, while for large $\alpha$ and/or $\mu<0$
we find $\partial N/\partial\alpha>0$. At fixed $\alpha$, the pair fraction decreases with $\mu$ (consistent with $\partial N/\partial\mu>0$).

The top curve in Fig.\ref{fraction}, representing $\mu=-1$, starts at $P=1$, roughly when $\alpha=1.5$.
Such a large value of $P$ is reminiscent of the BEC limit. The density vanishes here, then grows as $\alpha$ increases.
For $\mu=-1$, the pair fraction decreases with $\alpha$, except for a small kink, corresponding to the first order
$\xi=3\leftrightarrow\xi=2$ phase transition.

On the contrary, for $\mu=1$, $P$ grows with $\alpha$. As $\alpha\rightarrow0$,
$P$ becomes exponentially small, as is predicted by the BCS theory. After a sharp rise,
driven both by increasing $\alpha$ and decreasing $N$, the pair fraction levels out.

Each curve displays a kink, corresponding to the $\xi=3\leftrightarrow\xi=2$
phase transition. As $\alpha$ increases to the region $\xi=2$,
one row of vortices enters the elongated superfluid. This transition is accompanied
by density modulations.

To summarize we find that for $\mu>0$ and small $\alpha$ the system behaves analogously to the BCS limit,
while for $\mu<0$ and $\alpha\sim|\mu|$ the system behaves more like the BEC limit.
The density vanishes if $\mu<0$ and $\alpha\lesssim|\mu|$. For most of our parameter range,
we observe physics analogous to the crossover regime.

\textit{Vortex lattice --- }
With increasing $\alpha$, the number of Fourier components $\xi$ increases, and
the width in the $y$ direction grows. We illustrate the large $\alpha$ limit in Fig.\ref{lattice}
by calculating the density and the pair density of the state with $\mu=2,\alpha=65$ and $\xi=7$.
Only ``faint" vortices are seen in the density (left panel).
Unpaired fermions fill the vortex cores leading to very poor contrast. On the contrary,
one sees a clear stretched triangular lattice in the pair density (right panel).
The lattice spacing is $\sim2\pi\sqrt{2}\tilde\ell/K_0$ and the size of the vortex core is $\sim\tilde\ell$.
Note the dimensionless wave-number $K_0$ varies slightly with $\alpha$ but is of order $2$.
The vortex lattice is slightly deformed from a regular triangular lattice, but we expect this deformation to
disappear in the quasi-2D limit ($\alpha\rightarrow\infty$).

\textit{Observation --- }
Since the density depletion in the vortex core is highly suppressed,
directly imaging the vortices through phase contrast or absorption imaging would be challenging.
Coherent Bragg scattering of light may be a promising route for
increasing the sensitivity of such optical probes \cite{Sciamarella2001}.
One can also study the structures of pair density through
photoassociation \cite{photoassociation}, where the paired state
is transformed to a bound molecular state after illuminated with light.

\textit{Summary --- }
We have studied the two-component Fermi gases in elongated geometries. Truncating the BdG
equations to the lowest Landau level, we investigate the vortex structures
that emerge as the trap evolves from quasi-1D and quasi-2D.
We calculate the phase diagram and find instabilities towards pair density waves and vortex lattices.
We explore the structures of density and pair density, and calculate the pair fraction.
We hope our results can soon be explored in experiment.

\textit{Acknowledgements --- }
We thank S. S. Natu and S. Baur for carefully reading the manuscript.
R. W. is supported by CSC, the CAS, and the National Fundamental Research Program (under Grant
No. 2011CB921304). This material is based upon work supported by the National Science Foundation under Grant No. PHY-1068165.

\section{Supplemental materials}

\subsection{A -- Derivation of gap equation and number equation}
Here we analyze the special case where $\xi=1$, corresponding to a 1D model with translational invariance:
$\Delta_q=0$ unless $q=0$. Under these circumstances, Eq.(\ref{sime}) simplifies to

\begin{eqnarray}
\notag\mathcal{H}_0&=&\sum_{K,\sigma}(K^2-\mu)a_{K\sigma}^\dagger a_{K\sigma}\\
&+&\sum_K\left(\Delta_0^*f(0)+\Delta_0f^\dagger(0)\right)-|\Delta_0|^2/\beta.
\end{eqnarray}
where we have introduced the dimensionless Hamiltonian $\mathcal{H}_0=\notag H/{\cal E}$,
and $f(0)=\sum_Ke^{-\frac{1}{2}K^2}a_{-K\downarrow}a_{K\uparrow}$.

$\mathcal{H}_0$ can be diagonalized in terms of non-interacting Bogoliubov quasi-particle
operators $\xi_{K},\chi_{K}$ by the transformation
\begin{eqnarray}
\label{e1}
\left(\begin{array}{c}
    a_{K\uparrow} \\
    a^\dagger_{-K\downarrow}\\
    \end{array}\right)
    =\left(\begin{array}{cc}
    u_K & -v_K^* \\
    v_K & u_K^*\\
    \end{array}\right)
    \left(\begin{array}{c}
    \xi_K \\
    \chi_K^\dagger\\
    \end{array}\right)
\end{eqnarray}
yielding the diagonalized Hamiltonian,
\begin{eqnarray}
\label{e2}
\mathcal{H}_0=\sum_K\left(\epsilon_K(\xi_K^\dagger\xi_K+\chi_K^\dagger\chi_K)+\epsilon_0-\epsilon_K\right)-\frac{|\Delta_0|^2}{\beta}
\end{eqnarray}
where
\begin{eqnarray}
\label{e3}
u_K=\sqrt{\frac{\epsilon_K+\epsilon_0}{2\epsilon_K}},
v_K=\sqrt{\frac{\epsilon_K-\epsilon_0}{2\epsilon_K}}\\
\epsilon_{K}=\sqrt{\epsilon_0^2+|\Delta_0|^2e^{-K^2}},
\epsilon_0=K^2-\mu
\end{eqnarray}
We introduce the dimensionless energy $\mathcal{F}\equiv(2\sqrt{2}\pi\tilde\ell/L{\cal E})\langle GS|H|GS\rangle$,
where the ground state $|GS\rangle$ is annihilated by quasi-particle operators $\xi_{K},\chi_{K}$,
\begin{eqnarray}
\mathcal{F}=\int(\epsilon_0-\epsilon_{K})dK+|\Delta_{0}|^2/\alpha,
\end{eqnarray}
where we have taken the continuum limit $\sum_{K}\rightarrow(\sqrt{2}L/4\pi\tilde\ell)\int dK$.

Making $(1/|\Delta_0|)\partial\mathcal{F}/\partial|\Delta_0|=0$ yields the gap equation (\ref{gap}).
Letting $N=-(\sqrt{2}L/4\pi\tilde\ell)\partial\mathcal{F}/\partial\mu$ yields the number equation (\ref{number}).
These equations are further explored in the main text.

\subsection{B -- Numerical approach}
We here describe our numerical approach to solving the BdG equations in the general case where $\xi>1$.
Formally, Eq.(\ref{sime}) can be expressed in terms of non-interacting Bogoliubov quasi-particles
by a canonical transformation

\begin{eqnarray}
\label{trans}
\left(\begin{array}{c}
    a_{K_n\uparrow} \\
    a^\dagger_{K_n\downarrow}\\
    \end{array}\right)=\sum_{n'}
    \left(\begin{array}{cc}
    u_{n'n} & -v_{n'n}^* \\
    v_{n'n} & u_{n'n}^*\\
    \end{array}\right)\left(\begin{array}{c}
    \xi_{K_{n'}}\\
    \chi^\dagger_{K_{n'}}\\
    \end{array}\right),
\end{eqnarray}
where we have defined $K_n\equiv K-nK_0$, and $u_{n'n}\equiv u_{n'}(K_n),v_{n'n}\equiv v_{n'}(K_n)$.
The matrix elements $u_{n'n},v_{n'n}$ are governed by the following BdG equations,
\begin{eqnarray}
    \label{BdG}
    \epsilon_{K_n}\left(\begin{array}{c}
    v_{nn} \\
    u_{nn} \\
    \end{array}\right)
    =\sum_{n'}
    \left(\begin{array}{cc}
    -\varepsilon_{n'}\delta_{nn'} & (\Delta_{n'}^{n})^* \\
    \Delta_{n'}^{n} & \varepsilon_{n'}\delta_{nn'}\\
    \end{array}\right)
    \left(\begin{array}{c}
    v_{nn'} \\
    u_{nn'} \\
    \end{array}\right)
\end{eqnarray}
where $\epsilon_{K_n}$ is the dimensionless excitation energy of Bogoliubov quasi-particles,
and $\varepsilon_{n}=K_n^2-\mu$, $\Delta_{n'}^{n}=\Delta_{|n|K_0}e^{-\frac{1}{8}(2K_{n'}-nK_0)^2}$,
and $\delta_{nn'}$ is the $\delta$-function. In terms of the Bogoliubov operators the Hamiltonian is diagonal,
\begin{eqnarray}
\label{diag}
\notag H/{\cal E}&=&\sum_n\biggl[\sum_{K=-K_0/2}^{K_0/2}(\varepsilon_{n}-\epsilon_{K_n})-|\Delta_{|n|K_0}|^2/\beta\\
&+&\sum_{K=-K_0/2}^{K_0/2}\biggl(\epsilon_{K_n}(\xi_{K_n}^\dagger\xi_{K_n}+\chi_{K_n}^\dagger\chi_{K_n})\biggr)\biggr].
\end{eqnarray}
The dimensionless ground state energy $\mathcal{F}=(2\sqrt{2}\pi\tilde\ell/L{\cal E})\langle GS|H|GS\rangle$ can be written as
\begin{eqnarray}
\label{variational}
\mathcal{F}=\sum_n\biggl(\int_{-K_0/2}^{K_0/2}(\varepsilon_{n}-\epsilon_{K_n})dK
+|\Delta_{|n|K_0}|^2/\alpha\biggr).
\end{eqnarray}
For a given {$\{\mu,K_0,\Delta_{|n|K_0}\}$}, we truncate Eq.(\ref{BdG}), and use standard
linear algebra packages to extract $\epsilon_{K_n}$. This effectively gives us
$\mathcal{F}$ as a function of $\{\mu,\alpha,K_0,\Delta_{|n|K_0}\}$.
This $\mathcal{F}$ is a variational upperbound on the true ground state energy.
We fix $\{\mu,\alpha\}$ and numerically minimize $\mathcal{F}$, varying $\{K_0,\Delta_{|n|K_0}\}$,
using a quasi-Newton algorithm.
We restrict the sum over $n$ in Eq.(\ref{variational}) to $-\zeta\leq n\leq\zeta$.
We find for the parameters studied, our results are independent of $\zeta$ if $\zeta\geq6$.

\subsection{C -- Linear stability analysis}
Here we find the $\xi=1$ to $\xi=3$ phase boundary through a linear stability
analysis. We take $\Delta_0>0$, and assume $\Delta_{K_0}=\Delta_{-K_0}=i\delta$ is small.
We have chosen this factor of $i$, as the unstable direction will then yield real $\delta$.
We will calculate $D=\partial^2\mathcal{F}/\partial\delta^2|_{\delta=0}$.
For small $\alpha$ the curvature $D$ is positive and the state with $\delta=0$ is stable.
We find the instability by seeking the point with when $D=0$.

Within our ansatz for $\Delta_{|n|K_0}$, the mean field Hamiltonian is
\begin{eqnarray}
H/{\cal E}=\mathcal{H}_0+i\delta\Lambda-2\delta^2/\beta,
\end{eqnarray}
where
\begin{eqnarray}
\label{e4}
\Lambda&=&f^\dagger(K_0)+f^\dagger(-K_0)-f(K_0)-f(-K_0).
\end{eqnarray}

Making use of the Hellmann-Feynman theorem, the second derivative of $\mathcal{F}$ can be expressed as
\begin{eqnarray}
\notag\frac{\partial^2\mathcal{F}}{\partial\delta^2}&=&
\frac{2\sqrt{2}\pi\tilde\ell}{L}\frac{\partial^2}{\partial\delta^2}\langle GS|i\Lambda\delta-\frac{2\delta^2}{\beta}|GS\rangle\\
\notag&=&\frac{2\sqrt{2}\pi\tilde\ell}{L}\frac{\partial}{\partial\delta}\langle GS|i\Lambda-\frac{4\delta}{\beta}|GS\rangle\\
&=&-i\frac{4}{\alpha}\frac{\partial}{\partial\delta}\langle GS|\beta f^\dagger(K_0)|GS\rangle+\frac{4}{\alpha}.
\end{eqnarray}
Setting $D=\partial^2\mathcal{F}/\partial\delta^2|_{\delta=0}=0$, one finds that the points of instability is given by
\begin{eqnarray}
\label{e5}
-i=\beta\frac{\partial}{\partial\delta}\langle GS|f^\dagger(K_0)|GS\rangle.
\end{eqnarray}
Since the formal manipulations of perturbation theory are more transparent of finite temperature, it is convenient to
rewrite Eq.(\ref{e5}) as
\begin{eqnarray}
\label{e6}
\notag-i=\lim_{\mathcal{T}\rightarrow0}\beta\frac{\partial}{\partial\delta}
\frac{Tr(e^{-\mathcal{H}/\mathcal{T}}f^\dagger(K_0))}{Tr(e^{-\mathcal{H}/\mathcal{T}})}\\
=-i\beta\lim_{\mathcal{T}\rightarrow0}\frac{\int_0^{1/\mathcal{T}} Tr(e^{-\tau\mathcal{H}_0}\Lambda e^{(-1/\mathcal{T}+\tau)
\mathcal{H}_0}f^\dagger(K_0))d\tau}{Tr(e^{-\mathcal{H}_0/\mathcal{T}})}
\end{eqnarray}
where $\mathcal{T}$ is a formal parameter.

Substituting the results of Eq.(\ref{e2})-(\ref{e3}) to Eq.(\ref{e6}), we obtain
\begin{eqnarray}
\alpha\int\frac{(u_Ku_{K-K_0}+v_Kv_{K-K_0})^2e^{-\frac{1}{4}(K_0-2K)^2}}{\epsilon_{K-K_0}+\epsilon_K}dK=1.
\end{eqnarray}
This integral must be performed numerically, giving the second (right) black solid curve in Fig.\ref{pd}(a).


\begin{thebibliography}{99}

\bibitem{Zwierlein2005}
M. W. Zwierlein, J. R. Abo-Shaeer, A. Schirotzek, C. H. Schunck, and W. Ketterle,
Nature (London) \textbf{435}, 1047 (2005).

\bibitem{Sinha2005}
S. Sinha and G. V. Shlyapnikov, Phys. Rev. Lett. \textbf{94}, 150401 (2005).

\bibitem{Lin2009nature}
Y.-J. Lin, R. L. Compton, K. Jim\'{e}nez-Garc\'{i}a, J. V. Porto, and
I. B. Spielman, Nature (London) \textbf{462}, 628 (2009).



\bibitem{agterberg2008}
D. F. Agterberg, and, H. Tsunetsugu, Nat. Phys. \textbf{4}, 639 (2008).

\bibitem{LO1965}
A. I. Larkin, and Y. N. Ovchinnikov, Soviet Phys. JETP \textbf{20}, 762 (1965).



\bibitem{confinedvortices}
K. Kim, and D. R. Nelson, Phys. Rev. B, \textbf{64}, 054508 (2001);
W. Hofstetter, I. Affleck, D. Nelson, and U. Schollw\"{o}ck,
Europhys. Lett. \textbf{66}, 178 (2004);
I. Affleck, W. Hofstetter, D. R. Nelson, and U. Schollw\"{o}ck,
J. Stat. Mech.: Theor. Exp. \textbf{2004}, P10003 (2004).

\bibitem{superconductingwire}
J. S. Langer, and Q. Amberaogar, Phys. Rev. \textbf{64}, 498 (1967).




\bibitem{FF1964}
P. Fulde, and R. A. Ferrell, Phys. Rev. \textbf{135}, A550 (1964).

\bibitem{abrikosov1957}
A. A. Abrikosov, Soviet Phys. JETP \textbf{5}, 1174 (1957).

\bibitem{regulation}
Strictly speaking, this coupling constant $g$ should be regulized
as $-1/g=m/4\pi\hbar^2a_s-1/V\sum_{\bf{k}}1/2\epsilon_{\bf{k}}$,
where $V$ is the system volume and $\epsilon_{\bf{k}}$
is the excitation energy. However, as long as $a_s$ is
small compared to the transverse confinement, this regularization
does not change the effective 1D model \cite{Olshanii1998}.

\bibitem{Zwerger2004}
J. N. Fuchs, A. Recati, and W. Zwerger, Phys. Rev. Lett. \textbf{93}, 090408 (2004).

\bibitem{dalibard2002}
P. Rosenbusch, D. S. Petrov, S. Sinha, F. Chevy, V. Bretin, Y. Castin,
G. Shlyapnikov, and J. Dalibard, Phys. Rev. Lett. \textbf{88}, 250403 (2002).



\bibitem{singlevortex}
A. Bulgac, and Y. Yu, Phys. Rev. Lett. \textbf{91}, 190404 (2003);
N. Nygaard, G. M. Bruun, C. W. Clark, and D. L. Feder,
\textit{ibid}. \textbf{90}, 210402 (2003);
M. Machida, and T. Koyama,
\textit{ibid}.  \textbf{94}, 140401 (2005);
R. Sensarma, M. Randeria, and T.-L. Ho,
\textit{ibid}. \textbf{96}, 090403 (2006);
M. Takahashi, T. Mizushima, M. Ichioka, and K. Machida,
\textit{ibid}. \textbf{97}, 180407 (2006);
Hui Hu, Xia-Ji Liu, and Peter D. Drummond,
\textit{ibid}. \textbf{98}, 060406 (2007);
N. Nygaard, G. M. Bruun, B. I. Schneider, C. W. Clark, and D. L. Feder,
Phys. Rev. A \textbf{69}, 053622 (2004);
M. Machida, Y. Ohashi, and T. Koyama
\textit{ibid}. \textbf{74}, 023621 (2006);
H. J. Warringa, and A. Sedrakian,
\textit{ibid}. \textbf{84}, 023609 (2011).



\bibitem{vortexlattice}
D. L. Feder, Phys. Rev. Lett. \textbf{93}, 200406 (2004);
G. Tonini, F. Werner, and Y. Castin, Eur. Phys. J. D \textbf{39}, 283 (2006);
A. Bulgac, Y.-L. Luo, P. Magierski, K. J. Roche, and Y. Yu, Science \textbf{332}, 1288 (2011).


\bibitem{LLLBdG}
H. Akera, A. H. MacDonald, S. M. Girvin, and M. R. Norman,
Phys. Rev. Lett. \textbf{67}, 2375 (1991);
G. M\"{o}ller and N. R. Cooper,
\textit{ibid}. \textbf{99}, 190409 (2007);
H. Zhai, R. O. Umucal{\i}lar, and M. \"{O}. Oktel,
\textit{ibid}. \textbf{104}, 145301 (2010).


\bibitem{experimentLLL}
V. Schweikhard, I. Coddington, P. Engels, V. P. Mogendorff, and E. A. Cornell,
Phys. Rev. Lett. \textbf{92}, 040404 (2004);
V. Bretin, S. Stock, Y. Seurin, and J. Dalibard,
\textit{ibid}. \textbf{92}, 050403 (2004).



\bibitem{Aidelsburger2011}
M. Aidelsburger, M. Atala, S. Nascimb\`{e}ne, S. Trotzky,
Y.-A. Chen, and I. Bloch, Phys. Rev. Lett. \textbf{107}, 255301 (2011).


\bibitem{Salpeter1951}
E. E. Salpeter, Phys. Rev. \textbf{84}, 1226 (1951).


\bibitem{Olshanii1998}
M. Olshanii, Phys. Rev. Lett. \textbf{81}, 938 (1998).

\bibitem{Sciamarella2001}
D. Sciamarella, and Y. Pomeau, Journal of Low Temp. Phys. \textbf{123}, 35, (2001)

\bibitem{photoassociation}
K. M. Jones, E. Tiesinga, P. D. Lett, and P. S. Julienne,
Rev. Mod. Phys. \textbf{78}, 483 (2006);
G. B. Partridge, K. E. Strecker, R. I. Kamar, M. W. Jack, and R. G. Hulet,
Phys. Rev. Lett. \textbf{95}, 020404 (2005).

\bibitem{pair}
We define the number of fermions and paired fermions as
$N=\frac{\sqrt{2}L}{2\pi\tilde\ell}\int_{-K_0/2}^{K_0/2}
\sum_{n}\langle a_{K-nK_0\uparrow}^\dagger a_{K-nK_0\uparrow}\rangle dK$
and
$N_{\rm pair}=\frac{\sqrt{2}L}{4\pi\tilde\ell}\int_{-K_0/2}^{K_0/2}
\sum_{n,n'}|\langle a_{-K+nK_0\downarrow}a_{K-n'K_0\uparrow}\rangle|^2dK$.

\end{thebibliography}
\end{document}